\documentclass[12pt]{article}

\usepackage[labelfont=bf,labelsep=space]{caption}
\usepackage{times}
\usepackage{cite}
\usepackage{graphicx}
\usepackage{epstopdf}
\usepackage{dcolumn}
\usepackage{bm}
\usepackage{amsmath}
\usepackage{soul}

\usepackage{graphicx}

\newenvironment{sciabstract}{%
	\begin{quote} \bf}
	{\end{quote}}

\title{Visualizing heterogeneous dipole fields by terahertz light coupling in individual nano-junctions used in transmon qubits} 
\author
{R. H. J. Kim$^{1}$, J. M. Park$^{1}$, S. Haeuser$^{1,2}$, C. Huang$^{1,2}$,\\ D. Cheng$^{1,2}$, T. Koschny $^{1}$,  
	J. Oh$^{1}$, C. Kopas$^{3}$, H. Cansizoglu$^{3}$, K. Yadavalli$^{3}$,\\ J. Mutus$^{3}$, L. Zhou$^{1}$, L. Luo$^{1}$, M. Kramer$^{1}$, and J. Wang$^{1}$
	\\
	\\
	\normalsize{$^{1}$Ames Laboratory, Ames, IA 50011 USA}\\
	\normalsize{$^{2}$Department of Physics and Astronomy, Iowa State University, Ames, IA 50011 USA}\\
	\normalsize{$^{3}$Rigetti Computing, Berkeley, California, 94710, USA.} \\
}

\date{}

\begin{document} 
	
	
	\baselineskip24pt
	
	
	\maketitle

	
	\begin{sciabstract}
		The fundamental challenge underlying superconducting quantum computing is to characterize heterogeneity and disorder in the underlying quantum circuits. These nonuniform distributions often lead to local electric field concentration, charge scattering, dissipation and ultimately decoherence. 
		It is particularly challenging to probe deep sub-wavelength electric field distribution under electromagnetic wave coupling at individual nano-junctions and correlate 
		them with structural imperfections from interface and boundary, ubiquitous
		in Josephson junctions (JJ) used in transmon qubits. 
		A major obstacle lies in the fact that conventional microscopy tools are incapable of measuring {\em simultaneous} at nanometer and terahertz, ``nano-THz” scales, which often associate with frequency-dependent charge scattering in nano-junctions.  
		Here we directly visualize interface nano-dipole near-field distribution of individual Al/AlO$_{x}$/Al junctions used in transmon qubits. 
		Our THz nanoscope images show a remarkable asymmetry across the junction in electromagnetic wave-junction coupling response that manifests as ``hot" vs ``cold" cusp spatial electrical field structures and correlates with defected boundaries from the multi-angle deposition processes in JJ fabrication inside qubit devices.  
		The asymmetric nano-dipole electric field contrast also correlates with distinguishing, ``overshoot" frequency dependence that characterizes the charge scattering and dissipation at nanoscale, hidden in responses from topographic, structural imaging and spatially-averaged techniques.
		The real space mapping of junction dipole fields and THz charge scattering 
		can be extended to guide qubit nano-fabrication for ultimately optimizing qubit coherence times.  
	\end{sciabstract}
	$\newline$
	
The ability to identify heterogeneous electromagnetic field coupling into nano-junctions 
in quantum circuits is critical to understand and ultimately help guide materials selection and their processing to improve coherence in qubit devices \cite{QC1, QC2, QC3}. Disorder, structural asymmetry and interface imperfections are extremely difficult to avoid during complicated lithographic fabrication of the superconducting circuits and junctions.  
The existence of imperfections have been extensively studied using nano-structural instruments such as scanning and transmission electron microscopy. However, exactly how the structural heterogeneity affects electromagnetic wave coupling into the quantum circuits as well as electronic field distributions requires non-invasive electronic and photonic nano-THz probes that have not yet been used to map the nano-junctions used in any qubits.      
Our ability to visualize light-junction coupling at THz-nano scales will enable the determination of salient correlative features and identify previously unobserved error mechanisms, which significantly impact the optimization of fabrication procedures and understanding of decoherence/loss channels of materials and other qubit device components, e.g., Josephson junctions \cite{JJ1, JJ2}.

Combining few-cycle THz pulses\cite{yang2018,yang2019,liu2020,xyang2018,luo2014} with an atomic force microscope (AFM), time-domain THz scattering-type scanning near-field optical microscopy (THz-sSNOM)\cite{ribb2008,moon2014,agha2019,moon2019,liew2018,chen2003,zhang2018,wang2004,mais2019,rkim,xguo,stin2018,chen2020,yao2019,klar2017
} is an ideal way to capture local conductivity heterogeneity across nano-junctions at nm-THz scales. As illustrated in Fig. 1a, tip-scattered, time-domain THz spectroscopy measures both amplitude ($s_{n}$) and phase ($\phi_{n}$) of THz pulses (red arrow) on individual nano-junctions, shown in SEM image (center), which reveal broadband electrical polarizability and local ac charge transport. Although dipole fields in isotropic condensed media exhibit symmetric field and homogeneous electrodynamics, nano-dipoles that are polarized by electromagnetic couping at junction interfaces are affected by nanoscale inhomogeneities, such as sharp and broken boundaries, as illustrated in Fig. 1b.
Such asymmetric boundary conditions, along the upper and lower wings (bottom, Fig. 1a), are spatially correlated with the complicated multi-step deposition procedures used for fabricating Josephson junction structures in transmon qubit fabrication. As illustrated in Fig. 1a, the lower and upper L-shaped aluminum patterns are sequentially deposited on a Si substrate (gray, top left). Here, the AFM tip apex (light blue pyramid) serves as a nanoscale antenna for THz light that encodes the electronic heterogeneity, embedded in the near-field of nano-junction areas, to scattered THz light with distinguishing frequency dependence beyond the conventional topography and microscopy contrast.
Although near-field THz imaging has been prototyped in conventional semiconductors\cite{ribb2008,agha2019,liew2018}, semimetals\cite{rkim} and more recently aluminum resonators on silicon\cite{xguo}, 
this method has never been used to study nano-junctions, a critical component in superconduction qubit circuits, such as transmons.

\begin{figure}
	\begin{center}
		\includegraphics[width=160mm]{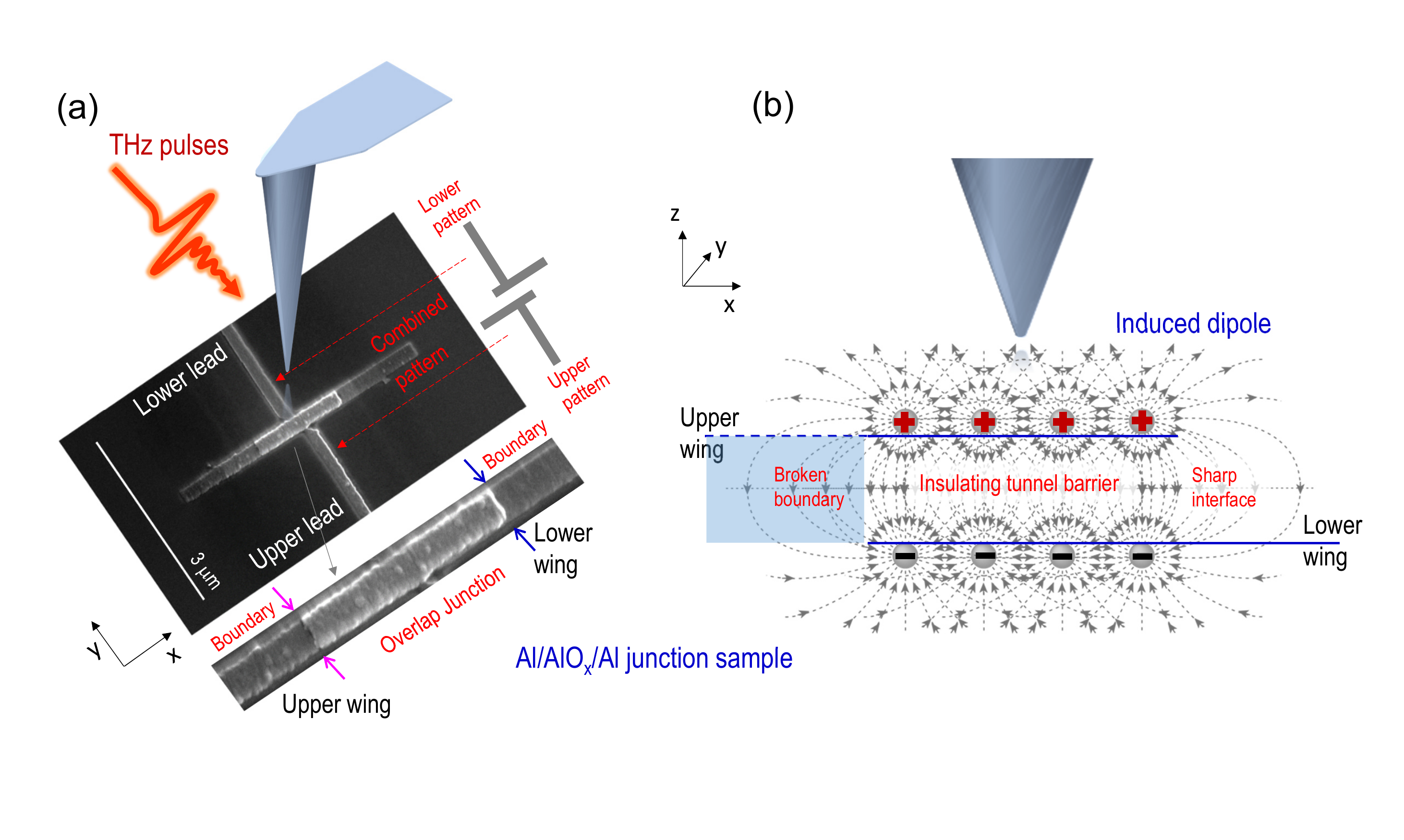}
	\end{center}
	\noindent {\textbf{Fig. 1. 	THz nano-imaging of Josephson-junction nano-dipole fields.} 
 (a) Schematic of the THz-sSNOM imaging method showing an incoming THz pulse to an AFM tip and an aluminum Josephson junction below it displayed as an SEM image. An enlarged image of the junction area is shown at the bottom. (b) Illustration of the formation of dipole fields at one instant of time as the incoming THz electric field induces free carriers in both sides of the metal lead to polarize the structure with opposite charges as in a capacitor.}
\end{figure}

In this letter, we discover asymmetric interface nano-dipole near field distributions and measure THz scattering spectra at sub-20 nm that visualize light coupling into individual nano-junctions. We clearly show that the junction boundaries and interfaces from lithographic procedures strongly affect the light-junction coupling that results in heterogeneous electric field distribution. 
The nano-electronic and photonic contrasts across junction boundaries manifest as distinguishing nm-THz structures, i.e., distinct variations in the THz-sSNOM amplitude that we denote as ``hot" and ``cold" cusp spatial structures, and a spectral ``overshoot response" where the near-field signal brightens towards the lower end of the THz frequency range. These newly discovered, nano-dipole electric field scattering contrast across nano-junctions are original and distinctly different from topographic steps that exhibit conventional ``inductive" spatial structures and featureless spectral shape.   

Our custom-made THz-sSNOM is based on a tapping-mode atomic force microscope (AFM) excited by ultrashort THz pulses. The AFM metallic tip acts as an antenna that receives far-field THz pulses, amplifies the near-field interaction through an enhancement of the THz field by tip resonances, and finally transmits and scatters the THz near-field to the far field for electro-optical (EO) sampling. Tip scattering signals $\mathrm{s}(t)$ are extracted by demodulating the backscattered radiation collected from the tip-sample system at the 2nd harmonics of the tip-tapping frequency.

Our near-field nano-imaging/spectroscopy experiment was performed on a set of JJ test devices from Rigetti Computing: Al/AlO$_{x}$/Al Josephson junctions fabricated on Si (100) substrates \cite{JJ2} using the nanofabrication procedure for its transmon qubits \cite{JJ1}. The junction comprises two L-shaped aluminum wire leads intersecting at a cross in the middle where a tunneling aluminum-oxide barrier of 1.3 nm separates the two metal layers as illustrated in Fig. 1a (gray pattern, top left). A scanning electron microscopy (SEM) image in Fig. 1a (center), shows the junction area of $\sim$1-to-2 micrometers across in size that is analyzed in our study. The JJ is defined by the region where an upper lead overlaps a lower lead separated by an insulating oxide layer shown at the center of the SEM image in Fig 1(a). The junction area is amplified (lower) to highlight the boundaries across the lower and upper wings. 
The JJs are fabricated in micro- to nano-scale area dimensions to reduce lossy channels from forming on the device interfaces and surfaces. Some degree of structural asymmetry and discontinuity are expected due to the geometry of double-angle shadow evaporation techniques commonly used to fabricate JJs.  The noninvasive THz nanoscopy technique allows us to probe the local electric field distributions in these areas and correlate them with structural features, such as dislocations, vacancies, and grain boundaries. 
The scattering spectra at THz-nano scales are critical to identifying unfavorable electric scattering and dissipation mechanisms that limit quality factors in individual nano-junctions. 
Once our THz pulse impinges on the sample in our near-field measurement, the highly enhanced electric field from the tip induces a series of oscillating dipole field patterns reminiscent to an ac-driven capacitor as illustrated in Fig. 1(b), as the tip is scanned across the surface. Opposite charges reside on the two different metallic layers at a given instant of time. Stray fields along the edges of the junction, where only the upper or the lower lead is found, are very sensitive to the local dielectric environment and can deviate from the uniform fields near the center; thus correlating nano-scale structural and electronic heterogeneous behavior at THz frequencies.

\begin{figure}
	\begin{center}
		\includegraphics[width=160mm]{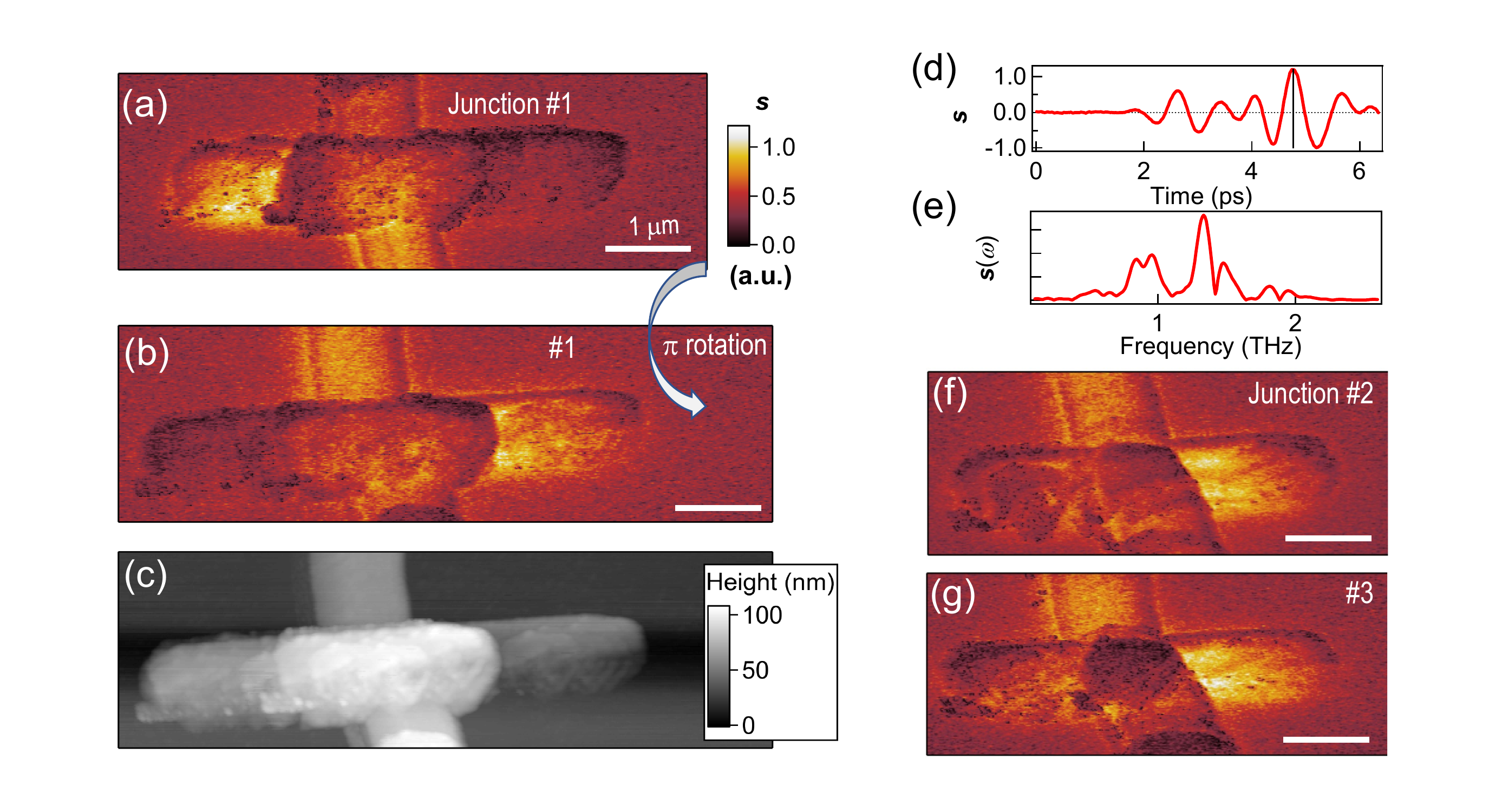}
	\end{center}
	\noindent {\textbf{Fig. 2. Representative THz near-field images of nano-dipole fields across Josephson junctions.} 
	 (a) Spatial mapping of the THz near-field scattering amplitude $s$ measured on a junction area in a 2D false color plot. (b) THz near-field image of the same area in (a) after rotating the sample 180 degrees in the sample plane. (c) A simultaneously acquired AFM topography image of the same region in (b). (d, e) A typical time-resolved THz near-field waveform measured on the sample and its Fourier transformed spectrum. (f, g) THz near-field contrast images of two additional Josephson junctions.}
\end{figure}

To this end, a direct visualization of the THz near-field electrodynamics through light-junction coupling demonstrates a dramatic asymmetric contrast across the junction area.  Figure 2(a) presents a typical scan on a Josephson junction showing a highly reflecting tip-scattered near-field amplitude from the lower aluminum junction wing and a darker contrast along the upper wing and lead structure. Figure 2(b) shows an identical scan of the region after rotating the sample 180 degrees in the $x$-$y$ plane. This confirms that the striking uneven brightness of the electric field scattering amplitude between the junction wing and lead wires is independent
on the direction of, and thereby, direct interaction with, the propagating field of the incident beam of THz light. It is rather related to the near-field coupling and nanoscale intrinsic junction properties. 
An AFM map taken simultaneously with our THz-sSNOM measurement, displayed in Fig. 2(c), further indicates that the asymmetric near-field signal across the nano-junction wings does not correlate with the topographic features of the junctions. The lower (upper) metal wing layer shows the brightest (darkest) THz scattered amplitude whereas the highest overlapping region at the center exhibit a moderate near-field intensity. Additional THz-sSNOM scans acquired on two other junction devices reveals the same asymmetric near field distribution of the scattered THz electric field with the nano-dipole field mostly concentrated toward the lower wing adjacent to the center junction. This conclusively demonstrates the discovered asymmetric nano-dipole field contrast is universal to the Josephson junctions fabricated using double-angle deposition techniques.

For a quantitative analysis of the THz near-field amplitude images, Figs. 3a-3b plot line scans running along the middle of the junction and across the single wire lower lead (inset, Fig. 3b) designed to connect the junction to the communication lines. The highly asymmetric feature in the nano-electronic contrast is easily discerned in the line profile of the THz near-field amplitude (thick red line) when compared with the height profile (thin gray line) measured simultaneously as shown in Fig. 3(a). We observed two distinct different contrasts. First, a topographic crosstalk leads a conventional THz near-field contrast that edge effects occur at sharp material boundaries near to the silicon substrate. This effect is most obvious at the edges of the wire leads as shown in Fig. 3(b). There is a sharp inductive spatial variation of the near-field signal as the AFM tip approaches the aluminum-substrate interface. At this lower edge, the AFM tip increases the total area of interaction with the sample, and thus increases the total near field signals. Consequently, the signal suddenly drops when the tip approachs the top of the edge and couples weakly with the sample structure. Second, the nano-dipole electric field scattering contrast sets in across nano-junctions as elaborated in detail below.  

\begin{figure}
	\begin{center}
		\includegraphics[width=160mm]{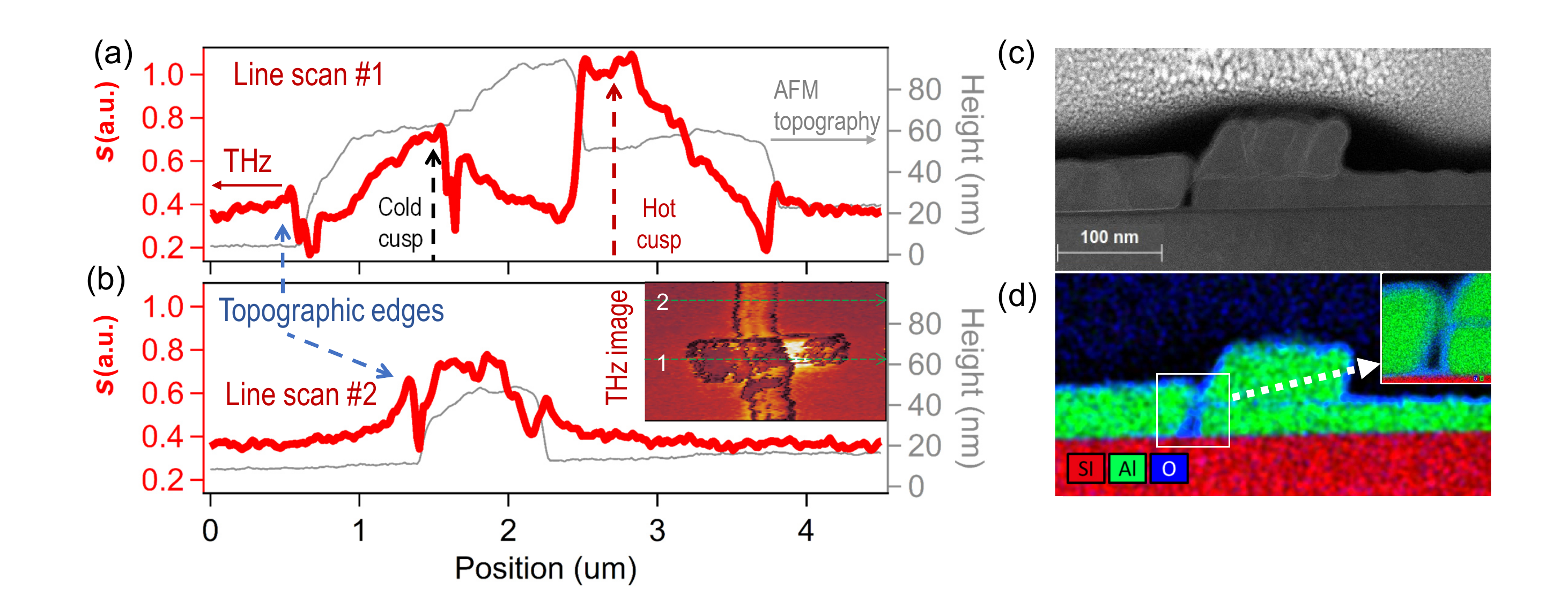}
	\end{center}
	\noindent {\textbf{Fig. 3. Nano-electronic/photonic vs. topographic profiles that show lithographic fingerprints of Josephson Junction fabrication.} 
	Line cuts of the near-field amplitude (red) and topography (gray) (a) along the direction from the upper wing to the lower wing across the center of the junction and (b) across the lower wire lead. (c) Cross-section high-angle annular dark-field STEM  image of a Josephson junction and the corresponding EDS elemental mapping (d) that shows the disconnection of the upper wing to the center piece of the junction.}
\end{figure}

A drastically different contrast appears near the junction interface, in comparison with the expected topographic near-field contrast around the low lead edges of the device (Fig. 3b). As shown in Fig. 3a (thick red line), the THz scattered amplitude increases as the tip moves toward the midpoint from both sides; however, the side associated with the upper aluminum wing displays a sudden drop in the amplitude as it meets the junction. Immediately next to the local maximum of the scattered signal of this upper lead, which we call the 'cold cusp', a deep trench is visible that most likely indicates a discontinuity or a break in the metallic structure. The deep trench is absent in the lower wing of the junction interface. 
At the very top of the junction, the near-field amplitude returns to the signal level that is similar to silicon. Most intriguingly, the THz amplitude reaches its maximum and forms a high plateau near the lower lead which we designate as the 'hot cusp'. 
Such spatial evolution of the 'hot' and 'cold' cusp features consistent with different local dielectric environment and nano-dipole electric fields. 
To correlate the local electronic response with structural one, a cross-section image was obtained along the center of the nano-junction by employing scanning transmission electron microscopy (STEM) and energy dispersive X-ray spectroscopy (EDS) elemental mapping, as shown in Fig. 3(c) and Fig. 3(d) respectively. 
We can see a clear separation along the upper lead that corresponds to the sharp near-field signal drop in the 'cold cusp' region. The EDS elemental mapping confirms the gap is partially filled with oxides and forms on one side of the nano-junction due to the possible shadowing from the step of the lower lead's edges during the deposition process of the upper leads. 
These distinguishable hot and cold near-field intensity differences between the lower and upper aluminum layers demonstrates that the THz near-field scattering amplitude provides valuable insights on possible spatial variations in the continuity between the two metallic layers. 
Note that the THz-sSNOM images are obtained in a nondestructive way while other mappings are obtained by destructively cutting the nano-junctions.

\begin{figure}
	\begin{center}
		\includegraphics[width=160mm]{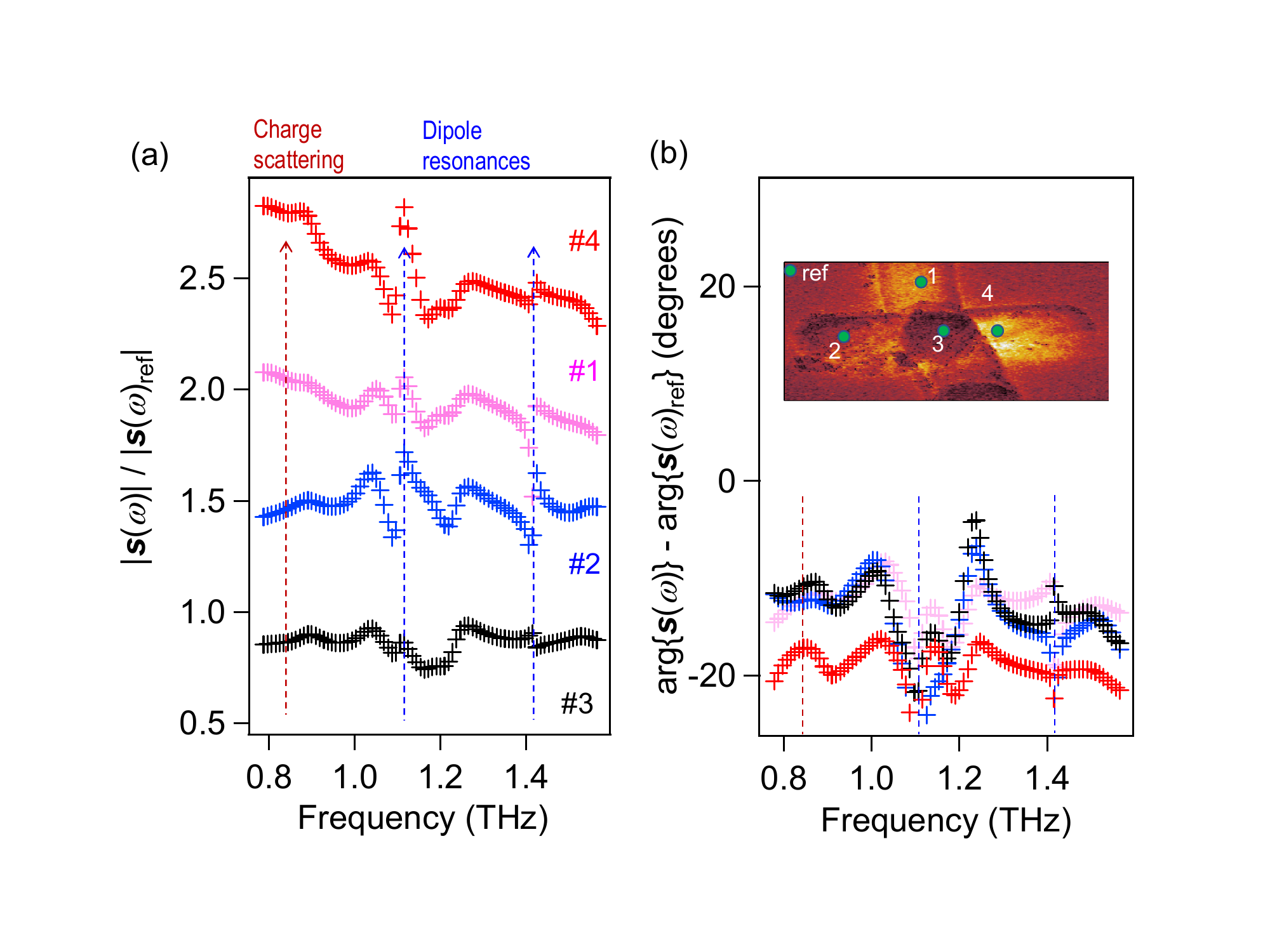}
	\end{center}
	\noindent {\textbf{Fig. 4. Heterogeneous electrodynamics in a single nano-junction.} 
	(a) Near-field spectral amplitude $|s(\omega)|$ and (b) phase arg$\{s(\omega)\}$ acquired by a Fourier transform of the time-domain signal as shown for example in Fig. 2(d). The results are normalized by another Fourier-transformed trace taken at the silicon substrate. Spectrums are taken at positions indicated in the inset of (b).}
\end{figure}

We further conducted THz nano-spectroscopy on different positions of the junction to examine the electronic scattering and dissipation channels implied in the observed heterogenous electrodynamics. 
By choosing four positions, as indicated in the inset of Fig. 4(b), we obtain the time dependent electric field waveform of the THz scattering response $s$, shown in Fig. 2(d). A fast Fourier transform was performed on the resulting time traces to yield a near-field amplitude $|s(\omega)|$, as shown in Fig. 2(e), and the phase arg$\{s(\omega)\}$ spectra. The resulting relative scattering amplitudes and phases are plotted in Fig. 4(a) and Fig. 4(b) respectively, with one spectrum recorded from the silicon substrate serving as a reference. The spectra includes the normal interruptions at 1.1 THz and 1.4 THz which arise from the dipole resonance of the AFM tip and cantilever assembly matching the THz wavelength. Additionally, the phase spectra include a slight shift from the insulating silicon substrate as is expected for the metallic components that form these junctions. Interestingly, we find that brighter 'hot cusp' regions are more than twice the magnitude in the near-field amplitude than that from darker 'cold cusp' portions of the junction. We also find the brighter 'hot cusp' regions exhibit a spectral slope curving upwards toward lower frequencies. As the entire metallic structure that connects to the small Josephson junction area amounts to the area of several hundreds of micrometers squared, it is likely that the light couples more efficiently with longer wavelengths of the impinging THz beam. Thus, the near-field scattered amplitude will be enhanced in the lower part of the THz frequency spectrum. 
As SEM images identified that grain sizes are smaller and the film being more granular for the lower layer aluminum, the higher signal contrast here determined from our THz sSNOM measurement suggests that the actual conductivity is governed by other factors, such as the sequential order of the substrate etching process and to the lithography steps used in the Josephson junction fabrication, while the granularity of the aluminum films is less of a contributing factor to the electric scattering.

In conclusion, we directly image THz light coupling into an individual nano-junction and discover deeply sub-wavelength heterogeneous near-field electrodynamics at nm scales.  
The near-field THz interrogation of a frequency-dependent effective polarizability provides a way of noninvasively identifying the junction interfaces and structural boundaries. The information obtained on the electric field distribution is highly complementary to destructive cross sectional characterization using STEM. Thus, the nondestructive time-domain THz scattering-type scanning near-field optical microscopy system shows promise in detecting possible variations in the thin-film fabrication processes hidden to surface imaging techniques.
The THz nanoscopy and amplitude-phase analysis are being further extended to cryogenic temperatures for capturing Cooper pair tunneling and circulation through Josephson junction devices
.

	\section*{Acknowledgments}
 This work was supported by the U.S. Department of Energy, Office of Science, National Quantum Information Science Research Centers, Superconducting Quantum Materials and Systems Center (SQMS) under the contract No. DE-AC02-07CH11359 (Josephson junction fabrication and characterizations, STEM characterizations). 
THz nano-microscopy/spectroscopy and modelling were supported by the Ames Laboratory, the US Department of Energy, Office of Science,
Basic Energy Sciences, Materials Science and Engineering Division under contract No. DEAC02-
07CH11358. THz sSNOM instrument was supported by the W.M. Keck Foundation for initial design and commission.


	\clearpage
	
	
\end{document}